\documentclass[applsci,article,submit,moreauthors,pdftex,10pt,a4paper]{mdpi}

\firstpage{1} 
\articlenumber{x}
\doinum{10.3390/------}
\pubvolume{xx}
\pubyear{2017}
\copyrightyear{2017}
\externaleditor{Academic Editor: name}
\history{Received: date; Accepted: date; Published: date}

 \theoremstyle{mdpi}
 \newcounter{thm}
 \newcounter{ex}
 \newcounter{re}
 \theoremstyle{mdpidefinition}

\Title{Toward Faultless Content-Based Playlists Generation for Instrumentals}

\Author{Yann Bayle $^{1,2}$*, Matthias Robine $^{1,2}$*, and Pierre Hanna $^{1,2}$*}
\AuthorNames{Yann Bayle, Matthias Robine and Pierre Hanna}

\address{%
$^{1}$ \quad Univ. Bordeaux, LaBRI, UMR 5800, F-33400 Talence, France\\
$^{2}$ \quad CNRS, LaBRI, UMR 5800, F-33400 Talence, France}

\corres{Correspondence: yann.bayle@u-bordeaux.fr, matthias.robine@u-bordeaux.fr, pierre.hanna@u-bordeaux.fr}

\abstract{
This study deals with content-based musical playlists generation focused on Songs and Instrumentals.
Automatic playlist generation relies on collaborative filtering and autotagging algorithms.
Autotagging can solve the cold start issue and popularity bias that are critical in music recommender systems.
However, autotagging remains to be improved and cannot generate satisfying music playlists.
In this paper, we suggest improvements toward better autotagging-generated playlists compared to state-of-the-art.
To assess our method, we focus on the Song and Instrumental tags.
Song and Instrumental are two objective and opposite tags that are under-studied compared to genres or moods, which are subjective and multi-modal tags.
In this paper, we consider an industrial real-world musical database that is unevenly distributed between Songs and Instrumentals and bigger than databases used in previous studies.
We set up three incremental experiments to enhance automatic playlist generation.
Our suggested approach generates an Instrumental playlist with up to three times less false positives than cutting edge methods.
Moreover, we provide a design of experiment framework to foster research on Songs and Instrumentals.
We give insight on how to improve further the quality of generated playlists and to extend our methods to other musical tags.
Furthermore, we provide the source code to guarantee reproducible research.
}


\begin{document}

\section{Introduction}

Playlists are becoming the main way of consuming music \citep{Song2012, Wikstrom2015, Choi2016, Nakano2016}.
This phenomenon is also confirmed on web streaming platforms, where playlists represent 40\% of musical streams as stated by De Gemini from Deezer\footnote{http://deezer.com, accessed on 27 September 2017} during the last MIDEM\footnote{http://musically.com/2016/06/05/music-curation-and-playlists-the-new-music-battleground-midem, accessed on 27 September 2017}.
Playlists also play a major role in other media like radios, personal devices such as laptops, smartphones \citep{Thalmann2016}, MP3 Players \citep{Nettamo2006}, and connected speakers.
Users can manually create their playlists, but a growing number of them listens to automatically generated playlists \citep{Uitdenbogerd2002} created by music recommender systems \citep{Yoshii2007,Schedl2015} that suggest tracks fitting the taste of each listener.

Such playlist generation implicitly requires selecting tracks with a common characteristic like genre or mood.
This equates to annotating tracks with meaningful information called tags \citep{Jaschke2007}.
A musical piece can gather one or multiple tags that can be comprehensible by common human listeners such as "happy", or not like "dynamic complexity" \citep{Streich2006,Laurier2007}.
A tag can also be related to the audio content, such as "rock" or "high tempo".
Moreover, editorial writers can provide tags like "summer hit" or "70s classic".
\citet{Turnbull2008} distinguish five methods to collect music tags.
Three of them require humans, e.g. social tagging websites \citep{Shardanand1995, Breese1998, Levy2007, Shepitsen2008} used by Last.fm\footnote{https://www.last.fm/, accessed on 27 September 2017}, music annotation games \citep{Law2007, Turnbull2007, Mandel2008}, and online polls \citep{Turnbull2008}.
The last two tagging methods are computer-based and include text mining web-documents \citep{Whitman2004, Knees2007} and audio content analysis \citep{Tzanetakis2002, Bertin-Mahieux2010, Prockup2015}.
Multiple drawbacks stand out when reviewing the different tagging methods.
Indeed, human labelling is time-consuming \citep{Kim2002, Skowronek2006} and prone to mistakes \citep{Sturm2013, Sturm2015}.
Furthermore, human labelling and text mining web-documents are limited by the ever-growing musical databases that increase by 4,000 new CDs by month \citep{Pachet1999} in western countries.
Hence, this amount of music cannot be labelled by humans and implies that some tracks cannot be recommended because they are not rated or tagged \citep{Eck2007, Li2007, Schafer2007, Schluter2015}.
This lack of labelling is a vicious circle in which unpopular musical pieces remain poorly labelled, whereas popular ones are more likely to be annotated on multiple criteria \citep{Eck2007} and therefore found in multiple playlists\footnote{http://www.billboard.com/biz/articles/news/digital-and-mobile/5944950/the-echo-nest-cto-brian-whitman-on-spotify-deal-man-vs, accessed on 27 September 2017}.
This phenomenon is known as the cold start issue or as the data sparsity problem \citep{Song2012}.
Text-mining web documents is tedious and error-prone, as it implies collecting and sorting redundant, contradictory, and semantic-based data from multiple sources.
Audio content-based tagging is faster than human labelling and solves the major problems of cold starts, popularity bias, and human-gathered tags \citep{Logan2002, Hoashi2003, Celma2005, Eck2007, Sordo2007, Turnbull2007, Mandel2008, Tingle2010}.
A makeshift solution combines the multiple tag-generating methods \citep{Bu2010} to produce robust tags and to process every track.
However, audio content analysis alone remains improvable for subjective and ambivalent tags such as the genre \citep{Hsu2016, Jeong2016, Lu2016, Oramas2016}.

In light of all these issues, a new paradigm is needed to rethink the classification problem and focus on a well-defined question\footnote{http://ejhumphrey.com/?p=302, accessed on 27 September 2017} that needs solving \citep{Sturm2016} to break the "glass ceiling" \citep{Wiggins2009} in Music Information Retrieval (MIR).
Indeed, setting up a problem with a precise definition will lead to better features and classification algorithms.
Certainly, cutting-edge algorithms are not suited for faultless playlist generation since they are built to balance precision and recall.
The presence of few wrong tracks in a playlist diminishes the trust of the user in the perceived service quality of a recommender system \citep{Chau2013} because users are more sensitive to negative than positive messages \citep{Yin2010}.
A faultless playlist based on a tag needs an algorithm that achieves perfect precision while maximizing recall.
It is possible to partially reach this aim by maximizing the precision and optimizing the corresponding recall, which is a different issue than optimizing the f-score.
A low recall is not a downside when considering the large amount of tracks available on audio streaming applications.
For example, Deezer provides more than 40 million tracks\footnote{https://www.deezer.com/features, accessed on 27 September 2017} in 2017.
Moreover, the maximum playlist size authorized on streaming platforms varies from 1,000\footnote{http://support.deezer.com/hc/en-gb/articles/201193652-Is-there-a-limit-to-the-amount-of-tracks-in-a-playlist-, accessed on 27 September 2017} for Deezer to 10,000\footnote{https://community.spotify.com/t5/Desktop-Linux-Windows-Web-Player/Maximum-songs-on-playlists/td-p/108021, accessed on 27 September 2017} for Spotify, while YouTube\footnote{https://developers.google.com/youtube/2.0/developers-guide-protocol-playlists?csw=1, accessed on 27 September 2017} and Google Play Music have a limit of 5,000 tracks per playlist.
However, there is a mean of 27 tracks in the private playlists of the users from Deezer with a standard variation of 70 tracks\footnote{Personal communication from Manuel Moussallam, Deezer R\&D team}.
Thus, it seems feasible to create tag-based playlists containing hundreds of tracks from large-scale musical databases.

In this article, we focus on improving audio content analysis to enhance playlist generation.
To do so, we perform Songs and Instrumentals Classification (SIC) in a musical database.
Songs and Instrumentals are well-defined, relatively objective, mutually exclusive, and always relevant \citep{Gouyon2014}.
We define a Song as a musical piece containing one or multiple singing voices either related to lyrics or onomatopoeias and that may or may not contain instrumentation.
Instrumental is thus defined as a musical piece that does not imply any sound directly or indirectly coming from the human voice.
An example of an indirect sound made by the human voice is the talking box effect audible in \textit{Rocky Mountain Way} from Joe Walsh.

People listen to instrumental music mostly for leisure.
However, we chose to focus on Instrumental detection in this study because Instrumentals are essential in therapy \citep{Rosenblatt2015} and learning enhancement methods \citep{Suarez2016, Zhao2016}.
Nevertheless, audio content analysis is currently limited by the distinction of singing voices from instruments that mimic voices.
Such distinction mistakes lead to plenty of Instrumental being labelled as Song.
Aerophones and fretless stringed instruments, for example, are known to produce similar pitch modulations as the human voice \citep{Rao2009, Panteli2017}.
This study focuses on improving Instrumental detection in musical databases because the current state-of-the-art algorithms are unable to generate a faultless playlist with the tag Instrumental \citep{Ghosal2013, Bayle2016}.
Moreover, precision and accuracy of SIC algorithms decline when faced with bigger musical databases \citep{Bayle2016, Bogdanov2016}.
The ability of these classification algorithms to generate faultless playlists is consequently discussed here.

In this paper, we define solutions to generate better Instrumental and Song playlists.
This is not a trivial task because Singing Voice Detection (SVD) algorithms cannot directly be used for SIC.
Indeed, SVD aims at detecting the presence of singing voice at the frame scale for one track, but related algorithms produce too many false positives \citep{Lehner2014}, especially when faced with Instrumentals.
Our work addresses this issue and the major contributions are:
\begin{itemize}[leftmargin=*,labelsep=4mm]
\item
The first review of SIC systems in the context of playlist generation.
\item
The first formal design of experiment of the SIC task.
\item
We show that the use of frame features outperforms the use of global track features in the case of SIC and thus diminishes the risk of an algorithm being a "Horse".
\item
A knowledge-based SIC algorithm ---easily explainable--- that can process large musical database whereas state-of-the-art algorithms cannot.
\item
A new track tagging method based on frame predictions that outperforms the Markov model in terms of accuracy and f-score.
\item
A demonstration that better playlists related to a tag can be generated when the autotagging algorithm focuses only on this tag.
\end{itemize}

As the major problem in MIR tasks concerns the lack of a big and clean labelled musical database \citep{Yoshii2007,Casey2008}, we thus detail in Section \ref{section_db} the use of \emph{SATIN} \citep{Bayle2017}, which is a persistent musical database.
This section also details the solution we use to guarantee reproducibility over \emph{SATIN} for our research code.
In Section \ref{section_stateoftheart} we describe the state-of-the-art methods in SIC and we detail their implementation in Section \ref{section_implementation}.
We then evaluate their performances and limitations in three experiments from Section \ref{section_expe1} to Section \ref{section_expe3}.
Section \ref{section_paradigm} settles the formalism for the new paradigm as described by \citep{Sturm2016} and compares our new proposed method to the state-of-the-art methods.
We finally discuss our results and perspectives in Section \ref{section_conclusion}.

\section{Musical database}
\label{section_db}

The musical database considered in this paper is twofold.
The first part of the musical database comprises 186 musical tracks evenly distributed between Songs and Instrumentals.
Tracks were chosen from previously existing musical databases.
This first part of our musical database is further referred as $D_p$.
All tracks are available for research purposes and are commonly used by the MIR community \citep{Ramona2008, Bittner2014, Lehner2014, Liutkus2014, Schluter2015,Schluter2016}.
$D_p$ includes tracks from the \emph{MedleyDB} database \citep{Bittner2014}, the \emph{ccMixter} database \citep{Liutkus2014}, and the \emph{Jamendo} database \citep{Ramona2008}.

\begin{itemize}[leftmargin=*,labelsep=4mm]
\item
The \emph{MedleyDB} database\footnote{http://medleydb.weebly.com, accessed on 27 September 2017} is a musical database of multi-track audio for music research proposed by \citet{Bittner2014}.
Forty-three tracks of \emph{MedleyDB} are used as Instrumentals in $D_p$.
\item 
The \emph{ccMixter} database contains 50 Songs compiled by \citet{Liutkus2014} and retrieved on \emph{ccMixter}\footnote{http://www.ccmixter.org, accessed on 27 September 2017}.
For each Song in the \emph{ccMixter} database, there is the corresponding Instrumental track.
These Instrumentals tracks are included in $D_p$.
\item
The \emph{Jamendo} database\footnote{http://www.mathieuramona.com/wp/data/jamendo, accessed on 27 September 2017} has been proposed by \citet{Ramona2008} and contains 93 Songs and the corresponding annotations at the frame scale concerning the presence of a singing voice.
These Songs have been retrieved from \emph{Jamendo} Music\footnote{https://www.jamendo.com, accessed on 27 September 2017}.
\end{itemize}
We chose tracks from the \emph{Jamendo} database because the MIR community already provided ground truths concerning the presence of a singing voice at the frame scale \citep{Ramona2008}.
These frame scale ground truths are indeed needed for the training process of the algorithm proposed in Section \ref{section_paradigm}.
There are only 93 Songs because producing corresponding frame scale ground truths is a tedious task, which is, to some extent, ill-defined \citep{Kim2002}.
We chose tracks from the \emph{MedleyDB} database because they are tagged as \textit{per se} Instrumentals, whereas we chose tracks from the \emph{ccMixter} database because they were meant to accompany a singing voice.
Choosing such different tracks helps to reflect the diversity of Instrumentals.

The second part of the musical database comes from the \emph{SATIN} \citep{Bayle2017} database and will be referred to as $D_s$.
$D_s$ is uneven and references 37,035 Songs and 4,456 Instrumentals, leading to a total of 41,491 tracks that are identified by their International Standard Recording Code (ISRC\footnote{http://isrc.ifpi.org/en, accessed on 27 September 2017}) provided by the International Federation of the Phonographic Industry (IFPI\footnote{http://www.ifpi.org/, accessed on 27 September 2017}).
These standard identifiers allow a unique identification of the different releases of a track over the years and across the interpretations from different artists.
The corresponding features of the tracks contained in \emph{SATIN} have been extracted for \citet{Bayle2017} by Simbals\footnote{http://www.simbals.com, accessed on 27 September 2017} and Deezer and are stored in \emph{SOFT1}.
To allow reproducibility, we provide the list of ISRC used for the following experiments along with our reproducible code on our GitHub account\footnote{https://github.com/ybayle/SMC2017, accessed on 27 September 2017\label{github}}.
The point of sharing the ISRC for each track is to facilitate result comparison between future studies and our own.

\section{State-of-the-art}
\label{section_stateoftheart}

As far as we know, only a few recent studies have been dedicated to SIC \citep{Ghosal2013, Hespanhol2013, Zhang2013, Gouyon2014, Bayle2016} compared to the extensive literature devoted to music genre recognition \citep{Sturm2014b}, for example.
The SIC task in a database must not be confused with the SVD task that tries to identify the presence of a singing voice at the frame scale for one track.
In this section, we describe existing algorithms for SIC and we benchmark them in the next section.

\subsection{Ghosal's Algorithm}

To segregate Songs and Instrumentals, \citet{Ghosal2013} extracted for each track the first thirteen Mel-Frequency Cepstral Coefficients (MFCC), excluding the $0^{th}$.
Indeed, akin to \citet{Zhang2013}, the authors posit that Songs differ from Instrumentals in the stable frequency peaks of the spectrogram visible in MFCC.
The authors then categorize an in-house database of 540 tracks evenly distributed with a classifier based on Random Sample and Consensus (RANSAC) \citep{Fischler1981,Ghosal2013}.
Their algorithm reaches an accuracy of 92.96\% for a 2-fold cross-validation classification task.
This algorithm will hereafter be denoted as GA.

\subsection{SVMBFF}

\citet{Gouyon2014} posit a variant of the algorithm from \citet{Ness2009}.
The seventeen low-level features extracted from each frame are normalized and consist of the zero crossing rate, the spectral centroid, the roll-off and flux, and the first thirteen MFCC.
A linear Support Vector Machine (SVM) classifier is trained to output probabilities for the mean and the standard deviation of the previous low-level features from which tags are selected.
The authors tested SVMBFF against three different musical databases comprising between 502 and 2,349 tracks.
The f-score of SVMBFF ranges from 0.89 to 0.95 for Songs across the three musical databases.
As for Instrumentals, the f-score is between 0.45 and 0.80.
The authors did not comment on this substantial variation and readers can foresee that the poor performance in Instrumental detection is not yet well understood.

\subsection{VQMM}

This approach has been proposed by \citet{Langlois2009} and enhanced by \citet{Gouyon2014}.
VQMM uses the YAAFE toolbox to compute the thirteen MFCC after the $0^{th}$ with an analysis frame of 93 ms and an overlap of 50\%.
VQMM then codes a signal using vector quantization (VQ) in a learned codebook.
Afterwards, it estimates conditional probabilities in first-order Markov models (MM).
The originality of this approach is found in the statistical language modelling.
The authors tested VQMM against three different musical databases comprising between 502 and 2,349 tracks.
The f-score of VQMM is comprised between 0.83 and 0.95 for Songs across the three musical databases.
The f-score for Instrumentals is between 0.54 and 0.66.
As for SVMBFF, the f-score of Instrumentals is lower than the f-score for Songs and depicts the difficulty to detect correctly Instrumentals, regardless of the musical database. 

\subsection{SRCAM}

\citet{Gouyon2014} used a variation of the sparse representation classification (SRC) \citep{Panagakis2009,Wright2009,Sturm2012,Sturm2012a} applied to auditory temporal modulation features (AM).
\citet{Gouyon2014} tested SRCAM against three different musical databases comprising between 502 and 2,349 tracks.
The f-score of SRCAM is comprised between 0.90 and 0.95 for Songs across the three musical databases.
The f-score for Instrumentals is between 0.57 and 0.80.
As for SVMBFF and VQMM, the f-score for Instrumentals is lower than the f-score for Songs.

\bigbreak

GA and SVMBFF use track scale features, whereas VQMM uses features at the frame scale.
The three algorithms use thirteen MFCC, as those peculiar features are well known to capture singing voice presence in tracks.
GA, SVMBFF, and VQMM are all tested under K-fold cross-validation on the same musical database.
In next section, we compare the performances of these three algorithms on the musical database $D_p$.

\section{Source code of the state-of-the-art for SIC}
\label{section_implementation}

This section describes the implementation we used to benchmark existing algorithms for SIC.
For all algorithms, the features proposed in \emph{SOFT1} were extracted and provided by Simbals and Deezer, thanks to the identifiers contained in \emph{SATIN}. 
More technical details about the classification process can be found on our previously mentioned GitHub repository.

\subsection{GA}

\citet{Ghosal2013} did not provide source code for reproducible research, so the YAAFE\footnote{http://yaafe.sourceforge.net, accessed on 27 September 2017} toolbox was used to extract the corresponding MFCC in this study.
The RANSAC algorithm provided by the Python package \emph{scikit-learn} \citep{scikit-learn} is used for classification.

\subsection{SVMBFF}

\citet{Gouyon2014} used the Marsyas framework\footnote{http://marsyas.info, accessed on 27 September 2017} to extract their features and to perform the classification, so we used the same framework along with the same parameters.

\subsection{VQMM}

The original implementation of VQMM made by \citet{Langlois2009} is freely available on their online repository\footnote{https://bitbucket.org/ThibaultLanglois/vqmm, accessed on 27 September 2017}.
We used this implementation with the same parameters that were used in their study.

\subsection{SRCAM}

SRCAM \citep{Gouyon2014} is dismissed as the source code is in Matlab.
Indeed, as tracks are stored on a remote industrial server, only algorithms for which the programming language is supported by our industrial partner can be computed. 
It would be interesting to implement SRCAM in Python or in C to assess its performance on $D_s$, but SRCAM displays similar results as SVMBFF on three different musical databases \citep{Gouyon2014}.

\section{Benchmark of existing algorithms for SIC}
\label{section_expe1}

In MIR, the aim of a classification task is to generate an algorithm capable of labelling each track of a musical database with meaningful tags.
Previous studies in SIC used musical databases containing between 502 and 2,349 unique tracks and performed a cross-validation with two to ten folds \citep{Ghosal2013, Hespanhol2013, Zhang2013, Gouyon2014, Bayle2016}.
This section introduces a similar experiment by benchmarking existing algorithms on a new musical database.
Table~\ref{table_1} displays the accuracy and the f-score of GA, SVMBFF, and VQMM with a 5-fold cross-validation classification task on $D_p$.

\begin{table}[H]
\caption{
Average $\pm$ standard deviation for accuracy and f-score for GA, SVMBFF, and VQMM with a 5-fold cross-validation classification task on the evenly balanced database $D_p$ of 186 tracks.
Bold numbers highlight the best results achieved for each metric.
}
\small
\centering
\begin{tabular}{ccc}
\toprule
Algorithm & Accuracy & F-score\\
\midrule
GA & $ 0.634 \pm 0.084 $ & $ 0.625 \pm 0.083 $\\
SVMBFF & $ 0.687 \pm 0.075 $ & $ 0.696 \pm 0.061$\\
VQMM & $ \boldsymbol{0.756} \pm 0.095 $ & $ \boldsymbol{0.753} \pm 0.099$\\
\bottomrule
\end{tabular}
\label{table_1}
\end{table}

The mean accuracy and f-score for the three algorithms do not differ significantly (one-way ANOVA, $F=2.600$, $p=0.120$).
The high variance, low accuracy, and the f-score of the three algorithms indicate that these algorithms are too dependent on the musical database and are not suitable for commercial applications.

K-fold cross-validation on the same musical database is regularly used as an accurate approximation of the performance of a classifier on different musical databases.
However, the size of the musical databases used in previous studies for SIC seems to be insufficient to assert the validity of any classification method \citep{Livshin2003, Guaus2009}.
Indeed, evaluating an algorithm on such small musical databases ---even with the use of K-fold cross-validation--- does not guarantee its generalization abilities because the included tracks might not necessarily be representative of all existing musical pieces \citep{Ng1997}.
K-fold cross-validation on small-sized musical databases is indeed prone to biases \citep{Herrera2003, Livshin2003, Bogdanov2011}, hence additional cross-database experiments are recommended in other scientific fields \citep{Chudacek2009, Bekios-Calfa2011, Llamedo2012, Erdogmus2014, Fernandez2015}.
Yet, creating a novel and large training set with corresponding ground truths consumes plenty of time and resources.
In fact, in the big data era, a small proportion of all existing tracks are reliably tagged in the musical databases of listeners or industrials, as can be seen on Last.fm or Pandora\footnote{https://www.pandora.com, accessed on 27 September 2017}, for example.
Thus, the numerous unlabelled tracks can only be classified with very few training data.
The precision of the classification reached in these conditions is uncertain.
The next section tackles this issue.

\section{Behaviour of the algorithms at scale}
\label{section_expe2}

This section compares the accuracy and the f-score of GA, SVMBFF, and VQMM in a cross-database validation experiment.
This experiment employs the test set $D_s$ that is 48 times bigger than the train set $D_p$.
This is a scale-up experiment compared to the number of tracks used in the previous experiment.
The reason for the use of a bigger test set is twofold.
Firstly, this behaviour mimics conditions in which there are more untagged than tagged data, which is common in the musical industry.
Secondly, existing classification algorithms for SIC cannot handle such an amount of musical data due to limitations of their own machine learning during the training process. 

The test set of 8,912 tracks is evenly distributed between Songs and Instrumentals. 
As there are fewer Instrumentals than Songs, all of them are used while eight successive random samples of Songs in $D_s$ are taken without replacement. 
In Table~\ref{table_2}, we compare the accuracy and f-score for GA, SVMBFF, and VQMM.

\begin{table}[H]
\caption{
Average $\pm$ standard deviation for accuracy and f-score for GA, SVMBFF, and VQMM.
The train set is constituted of the balanced database $D_p$ of 186 tracks.
The test set is successively constituted of eight evenly balanced sets of 8,912 tracks randomly chosen from the unbalanced database $D_s$ of 41,491 tracks.
Bold numbers highlight the best results achieved for each metric.
}
\small
\centering
\begin{tabular}{ccc}
\toprule
Algorithm & Accuracy & F-score\\
\midrule
GA & $ 0.623 \pm 0.017 $ & $ 0.604 \pm 0.014 $\\
SVMBFF & $ 0.566 \pm 0.021 $ & $ 0.542 \pm 0.027 $\\
VQMM & $ \boldsymbol{0.709} \pm 0.013 $ & $ \boldsymbol{0.707} \pm 0.012 $\\
\bottomrule
\end{tabular}
\label{table_2}
\end{table}

The accuracy and f-score of VQMM are higher than those of GA and SVMBFF, which may come from the use of local features by VQMM whereas GA and SVMBFF use track scale features.
Indeed, the accuracy and the f-score of GA, SVMBFF, and VQMM differ significantly (Posthoc Dunn test, $p<0.010$).
The accuracy of VQMM is respectively 0.086 (13.8\%) and 0.143 (25.3\%) higher than those of GA and SVMBFF.
The f-score of VQMM is respectively 0.103 (17.1\%) and 0.165 (30.4\%) higher than those of GA and SVMBFF.

Compared to the results of the first experiment in the same collection validation, the three algorithms have a lower accuracy: -0.011 (-1.7\%), -0.121 (-17.6\%), and -0.047 (-6.2\%), respectively for GA, SVMBFF, and VQMM.
The same trend is visible for the f-score with -0.021 (-3.4\%), -0.154 (-22.1\%), and -0.046 (-6.1\%), respectively for GA, SVMBFF, and VQMM.

The lower values of the accuracy and the f-score for the three algorithms in this experiment clearly depict the conjecture that same-database validation is not a suited experiment to assess the performances of an autotagging algorithm \citep{Herrera2003, Livshin2003, Guaus2009, Bogdanov2011}.
Moreover, the low values of the accuracy and the f-score of GA and SVMBFF in this untested database reveal that those algorithms might be "Horses" and might have overfit on the database proposed by their respective authors.
GA, SVMBFF, and VQMM are thus limited in accuracy and f-score when a bigger musical database is used, even if its size is far from reaching the 40 million tracks available \textit{via} Deezer.
It is highly probable that the accuracy and f-score of GA, SVMBFF, and VQMM will diminish further when faced with millions of tracks.

Furthermore, there is an uneven distribution of Songs and Instrumentals in personal and industrial musical databases.
Indeed, the salience of tracks containing singing voice in the recorded music industry is indubitable. 
Instrumentals represent 11 to 19\% of all tracks in musical databases\footnote{Personal communication from Manuel Moussallam, Deezer R\&D team}.
The next section investigates the possible differences in performance caused by this uneven distribution.

\section{Uneven class distribution}
\label{section_expe3}

This section evaluates the impact of a disequilibrium between Songs and Instrumentals on the precision, the recall, and the f-score of GA, SVMBFF, and VQMM.
It was not possible to perform a comparison between the existing algorithms dedicated to SIC using a K-fold cross-validation because the implementation of VQMM and SVMBFF cannot train on such a great amount of musical features and crashed when we tried to do so.
This section depicts a cross-database experiment with the 186 tracks of the balanced train set $D_p$ and the test set $D_s$ composed of 37,035 Songs (89\%) and 4,456 Instrumentals (11\%).
We compare in Table~\ref{table_3} the accuracy and the f-score of GA, SVMBFF, and VQMM.
To understand what is happening for the uneven distribution, we indicate which results are produced by a random classification algorithm further denoted RCA, i.e., where half of the musical database is randomly classified as Songs and the other half as Instrumentals.

\begin{table}[H]
\caption{
Average accuracy and f-score for GA, SVMBFF, and VQMM against a random classification algorithm denoted RCA.
The train set is constituted of the balanced database $D_p$ of 186 tracks.
The test set is constituted of the unbalanced database $D_s$ of 41,491 tracks composed of 37,035 Songs (89\%) and 4,456 Instrumentals (11\%).
Bold numbers highlight the best results achieved for each metric.
}
\small
\centering
\begin{tabular}{ccc}
\toprule
Algorithm & Accuracy & F-score\\
\midrule
GA & $ 0.769 $ & $ 0.795 $\\
RCA & $ 0.500 $ & $ 0.590 $\\
SVMBFF & $ 0.375 $ & $ 0.451 $\\
VQMM & $ \boldsymbol{0.784} $ & $ \boldsymbol{0.818} $\\

\bottomrule
\end{tabular}
\label{table_3}
\end{table}

VQMM, which uses frame scale features, has a higher accuracy and f-score than GA and SVMBFF, which use track scale features.
GA and VQMM perform better than RCA in terms of accuracy and f-score, contrary to SVMBFF.
The results of SVMBFF seem to depend on the context, i.e., on the musical database, because they display a lower global accuracy and f-score than RCA.
The poor performances of SVMBFF might be explained by the imbalance between Songs and Instrumentals.
As there is an uneven distribution between Instrumental and Songs in musical databases, we now analyse the precision, recall, and f-score for each class. 

\subsection{Results for Songs}

The Table~\ref{table_4} displays the precision and the recall for Songs detection for GA, SVMBFF, and VQMM against a random classification algorithm denoted RCA and \textit{via} the algorithm AllSong that classifies every track as Song.

\begin{table}[H]
\caption{
Song precision and Recall for the three algorithms defined in Section \ref{section_stateoftheart} against a random classification algorithm denoted RCA and \textit{via} an algorithm that classifies every track as Song denoted AllSong.
The train set is constituted of the balanced database $D_p$ of 186 tracks.
The test set is constituted of the unbalanced database $D_s$ of 41,491 tracks composed of 37,035 Songs (89\%) and 4,456 Instrumentals (11\%).
Bold numbers highlight the best results achieved for each metric.
}
\small
\centering
\begin{tabular}{cccc}
\toprule
Algorithm & Precision & Recall & F-score\\
\midrule
AllSong & $ 0.889 $ & $ \boldsymbol{1.000} $ & $ 0.941 $\\
GA & $ 0.908 $ & $ 0.824 $ & $ 0.864 $ \\
RCA & $ 0.889 $ & $ 0.500 $ & $ 0.640 $\\
SVMBFF & $ 0.932 $ & $ 0.324 $ & $ 0.480 $\\
VQMM & $ \boldsymbol{0.956} $ & $ 0.794 $ & $ \boldsymbol{0.967} $\\
\bottomrule
\end{tabular}
\label{table_4}
\end{table}

The precision for RCA and AllSong corresponds to the prevalence of the tag in the musical database.
RCA has a 50\% recall because half of the retrieved tracks is of interest, whereas AllSong has a recall of 100\%.
For GA, SVMBFF, and VQMM there is an increase in precision of respectively 0.02 (2.1\%), 0.04 (4.8\%), and 0.07 (7.5\%) compared to RCA and AllSong.

When all tracks are tagged as Song in a musical database it leads to a similar f-score than the state-of-the-art algorithm because Songs are in majority in such database.
Indeed, 100\% of recall is achieved by AllSong, which significantly increases the f-score.
The f-score is also increased by the high precision.
This precision corresponds to the prevalence of Songs, which are in majority in our musical database.
In sum, these results indicate that the best song playlist can be obtained by classifying every track of an uneven musical database as Song and that there is no need for a specific or complex algorithm.
We study in the next section the impact of such random classification on Instrumentals.

\subsection{Results for Instrumentals}

The Table~\ref{table_5} displays the precision and the recall for Instrumentals detection for GA, SVMBFF, and VQMM against RCA and \textit{via} the algorithm AllInstrumental that classifies every track as Instrumental.

\begin{table}[H]
\caption{
Instrumental precision and recall for the three algorithms defined in Section \ref{section_stateoftheart} against a random classification algorithm denoted RCA and \textit{via} an algorithm that classifies every track as Instrumental denoted AllInstrumental.
The train set is constituted of the balanced database $D_p$ of 186 tracks.
The test set is constituted of the unbalanced database $D_s$ of 41,491 tracks composed of 37,035 Songs (89\%) and 4,456 Instrumentals (11\%).
Bold numbers highlight the best results achieved for each metric.
}
\small
\centering
\begin{tabular}{cccc}
\toprule
Algorithm & Precision & Recall & F-score\\
\midrule
AllInstrumental & $ 0.110 $ & $ \boldsymbol{1.000} $ & $ 0.198 $\\
GA & $ 0.173 $ & $ 0.307 $ & $ 0.222 $\\
RCA & $ 0.110 $ & $ 0.500 $ & $ 0.181 $\\
SVMBFF & $ 0.125 $ & $ 0.803 $ & $ 0.216 $\\
VQMM & $ \boldsymbol{0.298} $ & $ 0.706 $ & $ \boldsymbol{0.419} $\\
\bottomrule
\end{tabular}
\label{table_5}
\end{table}

As with AllSong, the precision for RCA and AllInstrumental corresponds to the prevalence of the instrumental tag in $D_s$.
RCA has a 50\% recall because half of the retrieved tracks is of interest, whereas AllInstrumental has a recall of 100\%.
The precision of GA, SVMBFF, and VQMM is 0.06 (57.3\%), 0.02 (13.6\%), and 0.19 (170.9\%) higher respectively compared to RCA.
As for previous experiments, the better performance of VQMM over GA and SVMBFF might be imputable to the use of features at the frame scale. 
Even if the use of features at the frame scale by VQMM provides better performances than GA and SVMBFF, the precision remains very low for Instrumentals as VQMM only reaches 29.8\%.

In light of those results, guaranteeing faultless Instrumental playlists seems to be impossible with current algorithms.
Indeed, Instrumentals are not correctly detected in our musical database with state-of-the-art methods that reach, at best, a precision of 29.8\%.
As for the detection of Songs, classifying every track as a Song in our musical database produces a high precision that is only slightly improved by GA, SVMBFF, or VQMM.
A human listener might find inconspicuous the difference between a playlist generated by GA, SVMBFF, VQMM or by AllSong.
However, producing an Instrumental playlist remains a challenge.
The best Instrumental playlist feasible with GA, SVMBFF or VQMM contains at least 35 false positives ---i.e., Songs--- every 50 tracks, according to our experiments.
It is highly probable that listeners will notice it.
Thus, the precision of existing methods is not satisfactory enough to produce a faultless Instrumental playlist.
One might think a solution could be to select a different operating point on the receiver operating characteristic (ROC) curve.

\subsection{Results for different operating points}

\begin{figure}[H]
\centering
\includegraphics[width=10cm]{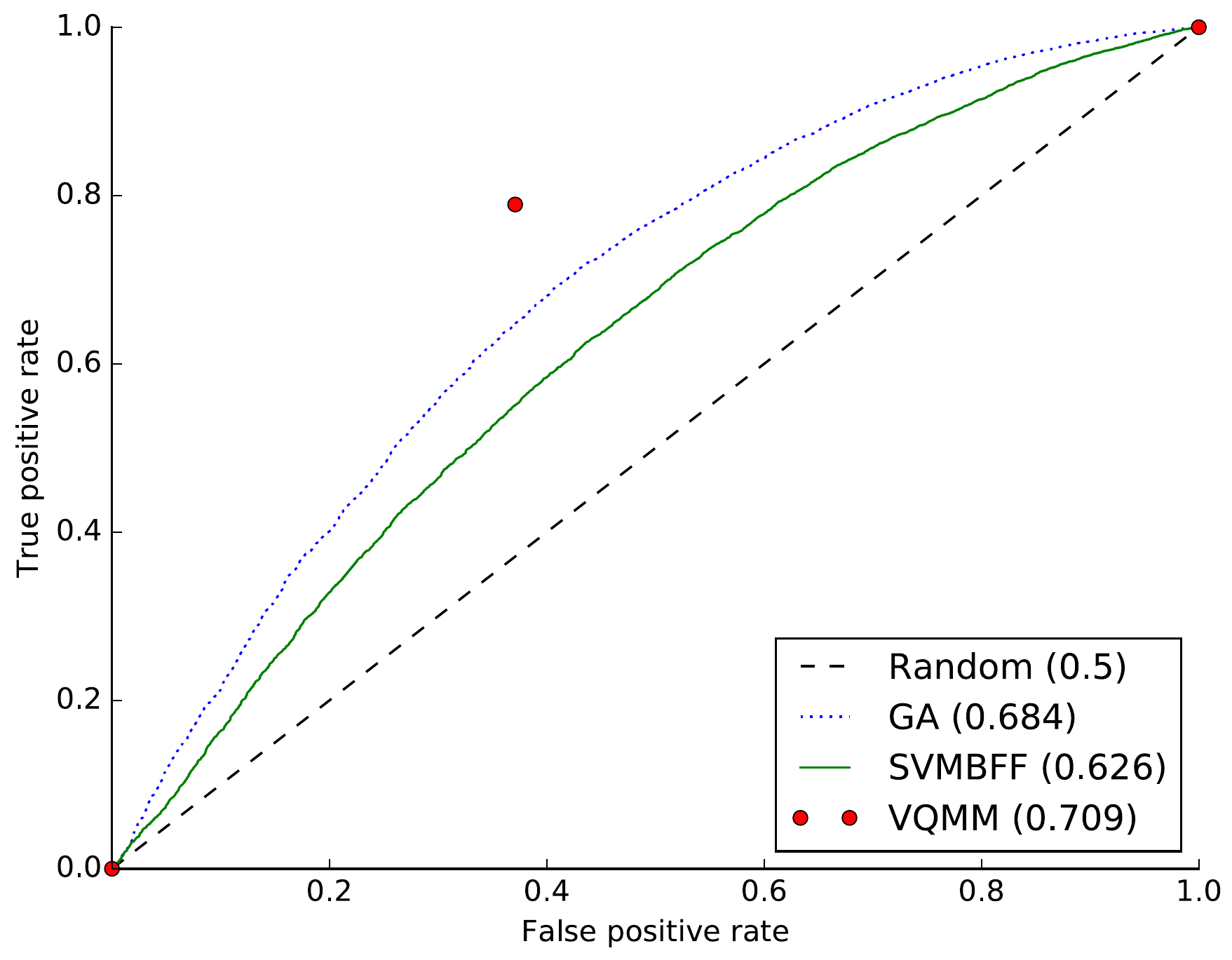}
\caption{
Receiver operating characteristic curve for the three algorithms defined in Section \ref{section_stateoftheart} along the area under the curve between brackets for the Songs.
The train set is constituted of the balanced database $D_p$ of 186 tracks.
The test set is constituted of the unbalanced database $D_s$ of 41,491 tracks composed of 37,035 Songs (89\%) and 4,456 Instrumentals (11\%).
}
\label{fig_ROC}
\end{figure}

Figure \ref{fig_ROC} shows the ROC curve for the three algorithms and the area under the curve (AUC) for the Songs.

The ROC curves of Figure \ref{fig_ROC} indicate that the only operating point for 100\% of true positive for GA, SVMBFF, and VQMM corresponds to 100\% of false positive.
Moreover, by design, there is a maximum of three operating points displayed by VQMM (Figure \ref{fig_ROC}).
Thus, a faultless playlist cannot be guaranteed by tuning the operating point of GA, SVMBFF, and VQMM.

\subsection{Class-weight alternative}

To guarantee a faultless playlist, another idea would be to tune algorithms by impacting the class weighting.
Indeed, we would guarantee 100\% precision even if the recall plummets.
Even if a recall of 1\% is reached on the 40 million tracks of Deezer, it provides a sufficient amount of tracks for generating 40 playlists fulfilling the maximum size authorized on streaming platforms.
Moreover, with such recall for the Instrumental tag, listeners can still apply another tag filter, such as "Jazz", to generate an Instrumental Jazz playlist, for example.

GA can be tuned, but not extensively enough to guarantee 100\% of precision because it uses RANSAC.
RANSAC is a regression algorithm robust to outliers and its configuration can only produce slight changes in performances, owing to its trade-off between accuracy and inliers.
VQMM can also be tuned, but the increase in performance is limited due to the generalization made by the Markov model.
SVMBFF can be tuned because class weights can be provided to SVM.
However, after trying different class weightings, the precision of SVMBFF only slightly varies, as the features used are not discriminating enough.

We also could have performed an N-fold cross-validation on $D_s$, but SVMBFF and VQMM cannot manage such an amount of musical data in the training phase. 

We thus propose using different features and algorithms to generate a better instrumental playlist than the ones possible with state-of-the-art algorithms.

\section{Toward better instrumental playlist}
\label{section_paradigm}

Experiments in previous sections indicate that GA, SVMBFF, and VQMM failed to generate a satisfactory enough Instrumental playlist out of an uneven and bigger musical database.
As previously mentioned, such a playlist requires the highest precision possible while optimizing the recall. 
GA, SVMBFF, and VQMM might be "Horses" \citep{Sturm2014a}, as they may not be addressing the problem they claim to solve.
Indeed, they are not dedicated to the detection of singing voice without lyrics such as onomatopoeias or the indistinct sound present in the song \textit{Crowd Chant} from Joe Satriani, for example.
To avoid similar mistakes, a proper goal \citep{Sturm2016} has to be clarified for SIC.
Indeed, a use case, a formal design of experiments (DOE) framework, and a feedback from the evaluation to system design are needed.

Our use case is composed of four elements: the music universe ($\Omega$), the music recording universe ($R_{\Omega}$), the description universe ($S_{\nu,A}$), and a success criterion.
$R_{\Omega}$ is composed of the polyphonic recording excerpts of the music in $\Omega$.
Songs and Instrumentals are the two classes of $S_{\nu,A}$.
The success criterion is reached when an Instrumental playlist without false positives is generated from autotagging.

Six treatments are applied.
Two are control treatments (Random Classification and the classification of every track as Instrumental), i.e. baselines.
Three treatments are state-of-the-art methods (GA, VQMM, and SVMBFF) and the last treatment is the proposed methodology.
The experimental units and the observational units are the entire collection of audio recordings.
As no cross-validation is processed, there is a unique treatment structure.
There are two responses model since our proposed algorithm has a two-stage process.
The first response model is binary because a track is either Instrumental or not.
The second response model is composed of the aggregate statistics (precision and recall).
The generated playlist is the treatment parameter.
The feedback is constituted of the number of Instrumentals in the final playlist. 
The experimental design of features and classifiers are detailed in the following section.
The treatment parameter is the generalization process made by our proposed algorithm, since this is the difference between the state-of-the-art algorithms and our proposed algorithm.
The materials in the DOE comes from the database \emph{SATIN} \citep{Bayle2017}.
We describe below the music universe ($\Omega$) ---i.e. \emph{SATIN}--- and its biases.
The biases in the database used in previous studies might have cause GA, VQMM, and SRCAM to overfit.
The biases in $\Omega$ have thus to be considered for the interpretation of the results.
\emph{SATIN} is a 41,491 semi-randomly sampled audio recordings out of 40M available on streaming platforms.
The sampling of tracks in \emph{SATIN} has been made in order to retrieve all the tracks that have a validated identifiers link between Deezer, Simbals, and Musixmatch.
\emph{SATIN} is representative in terms of genres and song/instrumental ratio.
\emph{SATIN} is biased towards the mainstream music as the tracks come from Deezer and Simbals.
The database does not include independent labels and artists that are available on SoundCloud, for example.
The tracks have been recorded in the last 30 years.
Finally, \emph{SATIN} is biased toward English artists because these represent more than one third of the database.

\subsection{Dedicated features for Instrumental detection}

\begin{figure}[H]
\centering
\includegraphics[width=10cm]{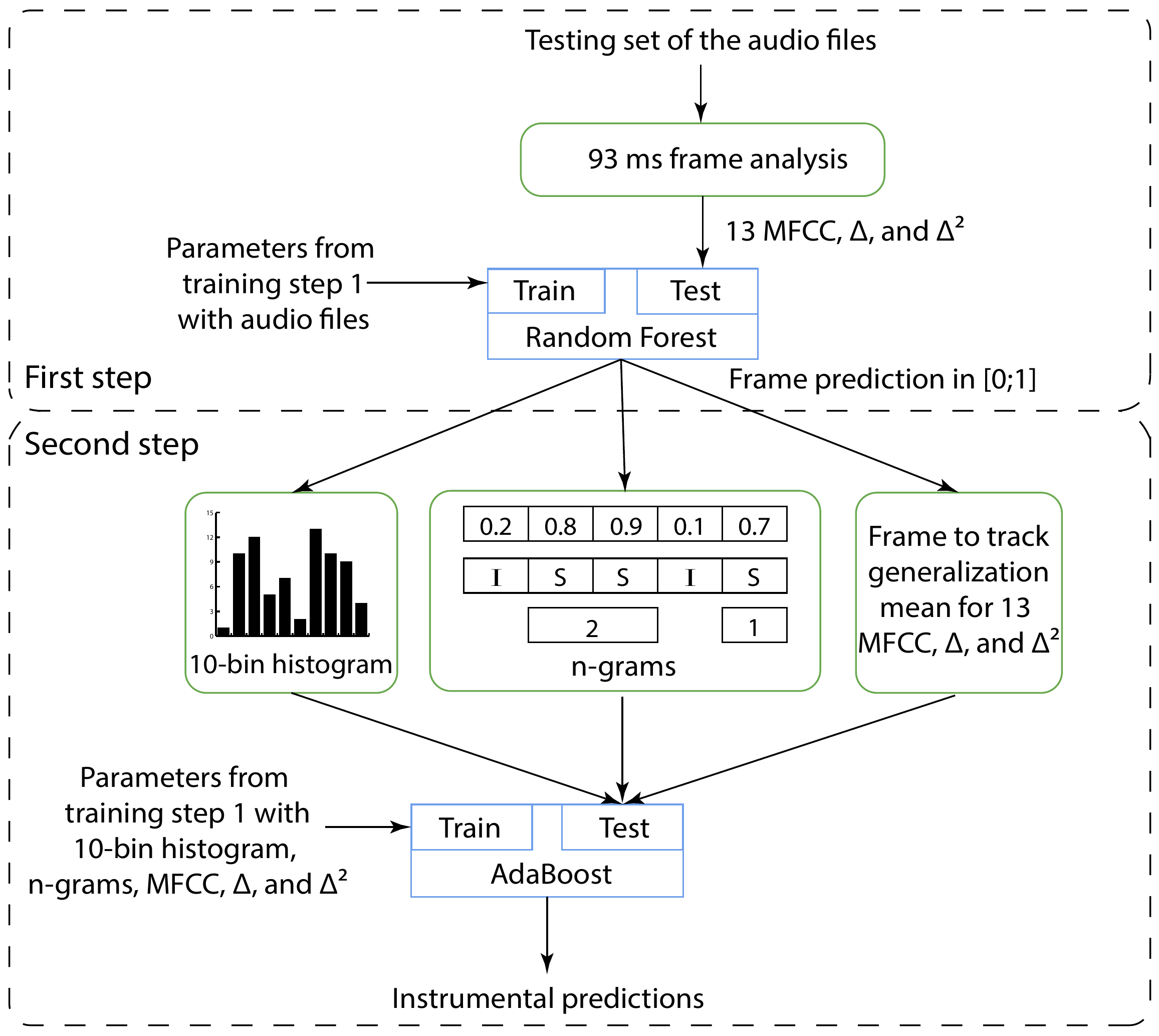}
\caption{
Schema detailing the algorithm for the detection of Instrumentals.
}
\label{fig_flow}
\end{figure}

The three experiments of this study show that using every feature at the frame scale increases more the performance than using features at the track scale.
In SVD, using frame features leads to Instrumentals misclassification, a high false positive rate, and indecision concerning the presence of singing voice at the frame scale.
However, for our task, using the classified frames together can enhance SIC and lead to better results at the track scale.
In order to use frame classification to detect Instrumentals, we propose a two-step algorithm.
The first step is similar to a regular SVD algorithm because it provides the probability that each frame contains singing voice or not.
In the second step, the algorithm uses the previously mentioned probabilities to classify each track as Song or Instrumental.
Figure \ref{fig_flow} details the underpinning mechanisms for the first step of Instrumental detection, which is a regular SVD method.

Our algorithm extracts the thirteen MFCC after the $0^{th}$ and the corresponding deltas and double deltas from each 93 ms frame of the tracks contained in $D_p$.
These features are then aligned with a frame ground truth made up by human annotators on the \emph{Jamendo} database \citep{Ramona2008}, which contains 93 Songs.
It is possible to have frame-precise alignments as the annotations provided by \citet{Ramona2008} are in forms of interval in which there is a singing voice or not. 
As for Instrumentals in $D_p$, all extracted features are associated with the tag Instrumental.
All these features and ground truths are then used to train a Random Forest classifier.
Afterwards, the Random Forest classifier outputs a vector of probability that indicates the likelihood of singing voice presence for each frame.

Now, each track has a probability vector corresponding to the singing voice presence likeliness for each frame.
The use of such soft annotations instead of binary ones has shown to improve the overall classification results \citep{Foucard2012}.
In the second step, the algorithm computes three sets of features for each track.
Two out of three are based on the previous probability vector.
The three sets of features generalize frame characteristics to produce features at the track scale.
The first set of features is a linear 10-bin histogram ranging from 0 to 1 by steps of 0.1 that represents the distribution of each probability vector.
Even if multiple frames are misclassified, the main trend of the histogram indicates that most frames are well classified.

\begin{figure}[H]
\centering
\includegraphics[width=10cm]{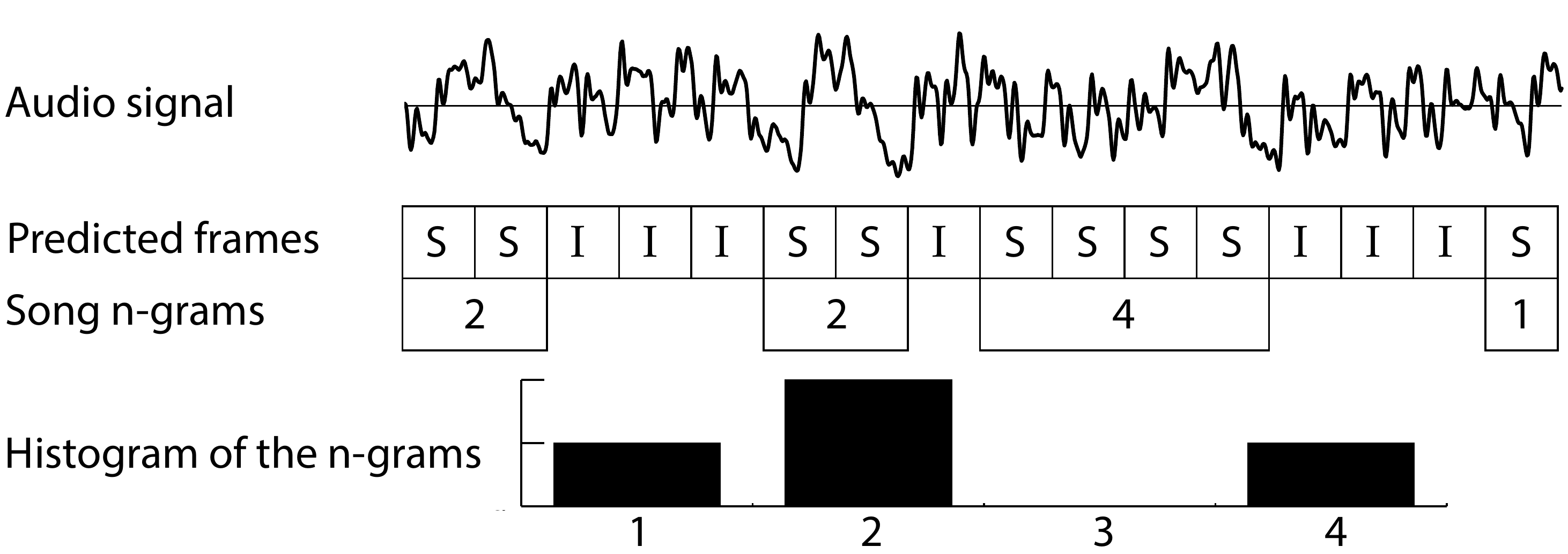}
\caption{
Detailed example for the n-gram construction.
}
\label{fig_ngram}
\end{figure}

Figure \ref{fig_ngram} details the construction of the second set of features ---named n-gram--- that uses the probability vector of singing voice presence.

These song n-grams are computed in two steps. 
In the first step, the algorithm counts the number of consecutive frames that were predicted to contain singing voice.
It then computes the corresponding normalized 30-bin histogram where n-grams greater than 30 are merged up with the last bin.
Indeed, chances are that an Instrumental will possess fewer consecutive frames classified as containing a singing voice than a Song. 
Consequently, an Instrumental can be distinguished from a Song by its low number of long consecutive predicted song frames.
By using this whole set of features against such an amount of musical data, we hope to keep "Horses" away \citep{Sturm2014,Sturm2014a}.
Indeed, we increase the probability that our algorithm is addressing the correct problem of distinguishing Instrumentals from Songs because of two reasons.
The first reason comes from the use of a sufficient amount of musical data that can reflects the diversity in music.
Indeed, our supervised algorithm can leverage instrumentals that contain violin to distinguish this amplitude modulation from the singing voice, for example.
This could not have been the case if the musical database was only constituted of rock music, for example.
The second reason comes from the features used that have been proven to detect the singing voice presence in multiple track modifications related to the pitch, the volume, and the speed \citep{Bayle2016}.
These kinds of musical data augmentation \citep{Schluter2015} are known to diminish the risk of overfitting \citep{Krizhevsky2012} and to improve the figures of merit in imbalanced class problems \citep{Chawla2009,Wong2016}, thus diminishing the risk of our algorithm being a "Horse".

Finally, the third and last set of features consists of the mean values for MFCC, deltas, and double deltas.

All these features are then used as training materials for an AdaBoost classifier, as described in the following section.

\subsection{Suited classification algorithm for Instrumental retrieval}

It is necessary to choose a machine learning algorithm that can focus on Instrumentals because these are not well detected and are in minority in musical databases. 
Thus, we choose to use boosting algorithms because they alter the weights of training examples to focus on the most intricate tracks. 
Boosting is preferred over Bagging, as the former aims to decrease bias and the latter aims to decrease variance.
In this particular applicative context of generating an Instrumental playlist from a big musical database, it is preferred to decrease the bias.
Among boosting algorithms, the AdaBoost classifier is known to perform well for the classification of minority tags \citep{Foucard2012} and music \citep{Bergstra2006}.
A decision tree is used as the base estimator in Adaboost.
The first reason for using decision trees lies in the logarithmic training curve displayed by decision trees and the second reason involves their better performances in the detection of the singing-voice by tree-based classifiers \citep{Lehner2014,Bayle2016}.
We use the AdaBoost implementation provided by the Python package \emph{scikit-learn} \citep{scikit-learn} to guarantee reproducibility.

\subsection{Evaluation of the performances of our algorithm}

This section evaluates the performances of the proposed algorithm in the same experiment as the one conducted in Section \ref{section_expe3}.
We remind the reader that we train our algorithm on the 186 tracks of $D_p$ and test it against the 41,941 tracks of $D_s$. 
Our algorithm reaches a global accuracy of $0.910$ and a global f-score of $0.883$.
Table~\ref{table_6} displays the precision and recall of our algorithm for Instrumentals classification and we display once again the previous corresponding results for AllInstrumental, GA, SMVBFF, and VQMM.

\begin{table}[H]
\caption{
Precision and recall of the new proposed algorithm.
The train set is constituted of the balanced database $D_p$ of 186 tracks.
The test set is constituted of the unbalanced database $D_s$ of 41,491 tracks composed of 37,035 Songs (89\%) and 4,456 Instrumentals (11\%).
The bold number highlights the best precision achieved.
}
\small
\centering
\begin{tabular}{cccc}
\toprule
Algorithm & Precision & Recall \\
\midrule
AllInstrumental & $ 0.110 $ & $ 1.000 $ \\
GA & $ 0.173 $ & $ 0.307 $ \\
RCA & $ 0.110 $ & $ 0.500 $ \\
SVMBFF & $ 0.125 $ & $ 0.803 $ \\
VQMM & $ 0.298 $ & $ 0.706 $ \\
\midrule
Proposed algorithm & \boldmath{$0.825$} & $ 0.200 $ \\
\bottomrule
\end{tabular}
\label{table_6}
\end{table}

As indicated in Table~\ref{table_6}, the main difference between our algorithm and GA, SVMBFF, and VQMM comes from the higher precision reached for Instrumental detection.
This precision of our algorithm is indeed 0.527 (276.8\%) higher than the best existing method ---i.e. VQMM--- and 0.715 (750.0\%) higher than RCA.
From a practical point of view, if GA, SVMBFF, and VQMM are used to build an Instrumental playlist, they can at best retrieve 30\% of true positive, i.e., Instrumentals, whereas our proposed method increases this number beyond 80\%, which is noteworthy for any listeners.
The high precision reached cannot be imputed to an over-fitting effect because the training set is 223 times smaller than the testing one.
The results from GA, SVMBFF and VQMM might have suffer from over-fitting because their experiment did imply a too restricted music universe ($\Omega$), in terms of size and representativeness of the tracks' origins.
Our algorithm brought the detection of Instrumentals closer to human-performance level than state-of-the-art algorithms.

When applying the same proposed algorithm to Songs instead of Instrumentals, our algorithm reaches a precision of 0.959 and a recall of 0.844 on Song detection, which is respectively 0.07 (7.9\%), and 34.4 (68.8\%) higher than RCA.
In this configuration, the global accuracy and f-score reached by our algorithm are respectively of 0.829 and 0.852.

\subsection{Limitations of our algorithm}

Just like for VQMM in Fig. \ref{fig_ROC}, we cannot tune our algorithm to guarantee 100\% of precision.
Our algorithm has only one operating point due to the use of the AdaBoost classifier.
We tried to use SVM and Random Forest classifiers --- which have multiple operating points --- but they cannot guarantee as much precision as AdaBoost did. 
Our algorithm in its current state performs better in Instrumental detection than state-of-the-art algorithms but it is still impossible to guarantee a faultless playlist.
As we aim to reduce the false positives to zero, the proposed classification algorithm seems to be limited by the set of features used.
A benchmark of SVD methods \citep{Lukashevich2007,Ramona2008,Regnier2009,Lehner2014,Leglaive2015,Lehner2015,Nwe2004,Schluter2015,Schluter2016} is needed to assess the impact of additional features on the precision and the recall when used with our generalization method.
Indeed, features such as the Vocal Variance \citep{Lehner2014}, the Voice Vibrato \citep{Regnier2009}, the Harmonic Attenuation \citep{Nwe2004} or the Auto-Regressive Moving Average filtering \citep{Lukashevich2007} have to be reviewed.

Apart from benchmarking features, a deep learning approach for SVD has been proposed \citep{Kereliuk2015, Leglaive2015, Lehner2015, Schluter2015, Lidy2016, Pons2016}.
However, deep learning is still a nascent and little understood approach in MIR\footnote{https://github.com/ybayle/awesome-deep-learning-music} and to the best of our knowledge no tuning of the operating point has been performed as it is intricate to analyse the inner layers \citep{Woods1997,Zhao2011}.
Furthermore, it is intricate to fit the whole spectrograms of full-length tracks of a given musical database into the memory of a GPU and thus it is intricate for a given deep learning model to train on full-length tracks on the SIC task.
Current deep learning approaches indeed require to fit into memory batches of tracks large enough ---usually 32 \citep{Miron2017a,Oramas2017a}--- to guarantee a good generalisation process.
For instance, neural network architecture for SVD algorithms like the one from \citet{Schluter2015} takes around 240MB in memory for 30 seconds spectrograms with 40 frequency bins for each track.
This architecture and batch size just fit in a high-end GPU with around 8GB of RAM.
To analyse full-length tracks of more than 4 minutes it would require to diminish the batch size below 4 thus decreasing harmfully the model generalization process.
This demonstration indicates that creating faultless instrumental playlist with a deep learning approach is not practically feasible now and currently the only solution toward better Instrumental playlists will require to enhance the input feature set of our algorithm.

\section{Conclusion}
\label{section_conclusion}

In this study, we propose solutions toward content-based driven generation of faultless Instrumental playlists.
Our new approach reaches a precision of 82.5\% for Instrumental detection, which is approximately three times better than state-of-the-art algorithms.
Moreover, this increase in precision is reached for a bigger musical database than the ones used in previous studies.

Our study provides five main contributions.
We provide the first review of SIC, which is in the applicative context of playlist generation ---in Section \ref{section_stateoftheart} to \ref{section_expe3}.
We show in Section \ref{section_paradigm} that the use of frame features outperforms the use of global track features in the case of SIC and thus diminishes the risk of an algorithm being a "Horse".
This improvement is magnified when frame ground truths are used alongside frame features, which is the key difference between our proposed algorithm and state-of-the-art algorithms.
Furthermore, our algorithm's implementation can process large musical databases whereas the current implementation of SVMBFF, SRCAM, and VQMM cannot.
Additionally, we propose in Section \ref{section_paradigm} a new track tagging method based on frame predictions that outperforms the Markov model in terms of accuracy and f-score.
Finally, we demonstrate that better playlists related to a tag can be generated when the autotagging algorithm focuses only on this tag.
This increase is accentuated when the tag is in minority, which is the case for most tags and especially here for Instrumentals.

\vspace{6pt} 

\supplementary{
The source code is available online at https://github.com/ybayle/SMC2017.
}

\acknowledgments{
The authors thank Thibault Langlois and Fabien Gouyon for their help reproducing VQMM and SVMBFF classification algorithms respectively.
The authors thank Manuel Moussallam from Deezer for the industrial acumen in music recommendations and fruitful discussions.
The authors thank Bob L. Sturm for his help formalizing the Songs and Instrumentals Classification task. 
The authors thank Jordi Pons for fruitful discussions on deep learning approaches.
The authors thank Fidji Berio and Kimberly Malcolm for insightful proofreading.
}

\authorcontributions{
All authors contributed equally to this work.
}

\conflictofinterests{
The authors declare no conflict of interest.
The industrial partners had no role in the design of the study; in the collection, analyses, or interpretation of data; in the writing of the manuscript, and in the decision to publish the results.
} 

\abbreviations{The following abbreviations are used in this manuscript:\\

\noindent ANOVA: ANalysis Of VAriance\\
AUC: Area Under the Curve\\
DOE: Design Of Experiments\\
GA: Ghosal's Algorithm\\
IFPI: International Federation of the Phonographic Industry\\
ISRC: International Standard Recording Code\\
MFCC: Mel-Frequency Cepstral Coefficients\\
MIR: Music Information Retrieval\\
RANSAC: Random Sample and Consensus\\
RCa: Random Classification Algorithm\\
ROC: Receiver Operating Characteristic\\
SATIN: Set of Audio Tags and Identifiers Normalized\\
SIC: Songs and Instrumental Classification\\
SRCAM: Sparse Representation Classification and Auditory temporal Modulation features\\
SVD: Singing Voice Detection\\
SVM: Support Vector Machine\\
SVMBFF: Support Vector Machine and Bags of Frames of Features\\
VQMM: Vector Quantization and Markov Models\\
}



\bibliographystyle{mdpi}


\bibliography{template}

\begin{thebibliography}{-------}
\providecommand{\natexlab}[1]{#1}

\bibitem[Song \em{et~al.}(2012)Song, Dixon, and Pearce]{Song2012}
Song, Y.; Dixon, S.; Pearce, M.
\newblock {A survey of music recommendation systems and future perspectives}.
\newblock  Proc. 9th Int. Symp. Comp. Music Model. Retrieval,  2012, pp.
  395--410.

\bibitem[Wikstr{\"{o}}m(2015)]{Wikstrom2015}
Wikstr{\"{o}}m, P.
\newblock {Will algorithmic playlist curation be the end of music stardom?}
\newblock {\em J. Bus. Anthrop.} {\bf 2015}, {\em 4},~278--284.

\bibitem[Choi \em{et~al.}(2016)Choi, Fazekas, and Sandler]{Choi2016}
Choi, K.; Fazekas, G.; Sandler, M.
\newblock {Towards playlist generation algorithms using RNNs trained on
  within-track transitions}.
\newblock  Work. Surprise Opposition Obstruction Adapt. Personalized Syst.,
  2016, pp. 1--4.

\bibitem[Nakano \em{et~al.}(2016)Nakano, Kato, Hamasaki, and Goto]{Nakano2016}
Nakano, T.; Kato, J.; Hamasaki, M.; Goto, M.
\newblock {PlaylistPlayer: an interface using multiple criteria to change the
  playback order of a music playlist}.
\newblock  Proc. 21st Int. Conf. Intell. User Interfaces. ACM,  2016, pp.
  186--190.

\bibitem[Thalmann \em{et~al.}(2016)Thalmann, {Perez Carillo}, Fazekas, Wiggins,
  and Sandler]{Thalmann2016}
Thalmann, F.S.; {Perez Carillo}, A.; Fazekas, G.; Wiggins, G.A.; Sandler, M.B.
\newblock {The Semantic Music Player: a smart mobile player based on
  ontological structures and analytical feature metadata}.
\newblock  Proc. 2nd Web Audio Conf.,  2016, pp. 1--6.

\bibitem[Nettamo \em{et~al.}(2006)Nettamo, Nirhamo, and
  H{\"{a}}kkil{\"{a}}]{Nettamo2006}
Nettamo, E.; Nirhamo, M.; H{\"{a}}kkil{\"{a}}, J.
\newblock {A cross-cultural study of mobile music: retrieval, management and
  consumption}.
\newblock  Proc. 18th Australia Conf. Comput. Human Interaction Design
  Activities Artefacts Environ.,  2006, pp. 87--94.

\bibitem[Uitdenbogerd and Schyndel(2002)]{Uitdenbogerd2002}
Uitdenbogerd, A.; Schyndel, R.
\newblock {A review of factors affecting music recommender success}.
\newblock  Proc. 3rd Int. Conf. Music Inform. Retrieval,  2002, pp. 204--208.

\bibitem[Yoshii \em{et~al.}(2007)Yoshii, Goto, Komatani, Ogata, and
  Okuno]{Yoshii2007}
Yoshii, K.; Goto, M.; Komatani, K.; Ogata, T.; Okuno, H.G.
\newblock {Improving efficiency and scalability of model-based music
  recommender system based on incremental training}.
\newblock  Proc. 8th Int. Conf. Music Inform. Retrieval,  2007, pp. 89--94.

\bibitem[Schedl \em{et~al.}(2015)Schedl, Knees, McFee, Bogdanov, and
  Kaminskas]{Schedl2015}
Schedl, M.; Knees, P.; McFee, B.; Bogdanov, D.; Kaminskas, M.
\newblock {Music recommender systems}. In {\em Recommender Systems Handbook};
  Ricci, F.; Rokach, L.; Shapira, B., Eds.; Springer US,  2015; chapter~13, pp.
  453--492.

\bibitem[J{\"{a}}schke \em{et~al.}(2007)J{\"{a}}schke, Marinho, Hotho,
  Schmidt-Thieme, and Stumme]{Jaschke2007}
J{\"{a}}schke, R.; Marinho, L.; Hotho, A.; Schmidt-Thieme, L.; Stumme, G.
\newblock {Tag recommendations in folksonomies}.
\newblock  Proc. 11th Euro. Conf. Princ. Practice Know. Disc. Databases,  2007,
  pp. 506--514.

\bibitem[Streich(2006)]{Streich2006}
Streich, S.
\newblock {Music complexity: a multi-faceted description of audio content}.
\newblock PhD thesis, Univ. Pompeu Fabra, Barcelona, Spain,  2006.

\bibitem[Laurier and Herrera(2007)]{Laurier2007}
Laurier, C.; Herrera, P.
\newblock {Audio music mood classification using support vector machine}.
\newblock  MIREX task on Audio Mood Classification,  2007, pp. 1--3.

\bibitem[Turnbull \em{et~al.}(2008)Turnbull, Barrington, Torres, and
  Lanckriet]{Turnbull2008}
Turnbull, D.; Barrington, L.; Torres, D.; Lanckriet, G.
\newblock {Semantic annotation and retrieval of music and sound effects}.
\newblock {\em {IEEE} Trans. Audio Speech Lang. Process.} {\bf 2008}, {\em
  16},~467--476.

\bibitem[Shardanand and Maes(1995)]{Shardanand1995}
Shardanand, U.; Maes, P.
\newblock {Social information filtering: algorithms for automating “word of
  mouth”}.
\newblock  Proc. Spec. Interest Group Comp. Human Interact. Conf. Human Factors
  in Comp. Syst.,  1995, pp. 210--217.

\bibitem[Breese \em{et~al.}(1998)Breese, Heckerman, and Kadie]{Breese1998}
Breese, J.S.; Heckerman, D.; Kadie, C.
\newblock {Empirical analysis of predictive algorithms for collaborative
  filtering}.
\newblock  Proc. 14th Conf. Uncertainty Artif. Intell.,  1998, pp. 43--52.

\bibitem[Levy and Sandler(2007)]{Levy2007}
Levy, M.; Sandler, M.B.
\newblock {A semantic space for music derived from social tags}.
\newblock  Proc. 8th Int. Conf. Music Inform. Retrieval,  2007, pp. 411--416.

\bibitem[Shepitsen \em{et~al.}(2008)Shepitsen, Gemmell, Mobasher, and
  Burke]{Shepitsen2008}
Shepitsen, A.; Gemmell, J.; Mobasher, B.; Burke, R.
\newblock {Personalized recommendation in social tagging systems using
  hierarchical clustering}.
\newblock  Proc. ACM 2nd Conf. Recomm. Syst.,  2008, pp. 259--266.

\bibitem[Law \em{et~al.}(2007)Law, von Ahn, Dannenberg, and Crawford]{Law2007}
Law, E.L.M.; von Ahn, L.; Dannenberg, R.B.; Crawford, M.
\newblock {Tagatune : a game for music and sound annotation}.
\newblock  Proc. 8th Int. Conf. Music Inform. Retrieval,  2007, pp. 361--364.

\bibitem[Turnbull \em{et~al.}(2007)Turnbull, Barrington, Torres, and
  Lanckriet]{Turnbull2007}
Turnbull, D.; Barrington, L.; Torres, D.; Lanckriet, G.
\newblock {Towards musical query-by-semantic-description using the CAL500 data
  set}.
\newblock  Proc. 30th Annu. Int. ACM SIGIR Conf. Res. Devl. Inf. Retr.,  2007,
  pp. 439--446.

\bibitem[Mandel and Ellis(2008)]{Mandel2008}
Mandel, M.I.; Ellis, D.P.W.
\newblock {Multiple-instance learning for music information retrieval.}
\newblock  Proc. 9th Int. Conf. Music Inform. Retrieval,  2008, pp. 577--582.

\bibitem[Whitman and Ellis(2004)]{Whitman2004}
Whitman, B.; Ellis, D.P.W.
\newblock {Automatic record reviews}.
\newblock  Proc. 5th Int. Conf. Music Inform. Retrieval,  2004, pp. 470--477.

\bibitem[Knees \em{et~al.}(2007)Knees, Pohle, Schedl, and Widmer]{Knees2007}
Knees, P.; Pohle, T.; Schedl, M.; Widmer, G.
\newblock {A music search engine built upon audio-based and web-based
  similarity measures}.
\newblock  Proc. 30th Annu. Int. ACM SIGIR Conf. Res. Devl. Inf. Retr.,  2007,
  pp. 447--454.

\bibitem[Tzanetakis and Cook(2002)]{Tzanetakis2002}
Tzanetakis, G.; Cook, P.
\newblock {Musical genre classification of audio signals}.
\newblock {\em {IEEE} Trans. Speech Audio Process.} {\bf 2002}, {\em
  10},~293--302.

\bibitem[Bertin-Mahieux \em{et~al.}(2010)Bertin-Mahieux, Eck, and
  Mandel]{Bertin-Mahieux2010}
Bertin-Mahieux, T.; Eck, D.; Mandel, M.I.
\newblock {Automatic Tagging of Audio: The State-of-the-Art}. In {\em Mach.
  Audition Prin. Algo. Syst.}; Wang, W., Ed.; Information Science Reference,
  IGI Global,  2010; chapter~14, pp. 334--352.

\bibitem[Prockup \em{et~al.}(2015)Prockup, Ehmann, Gouyon, Schmidt, Celma, and
  Kim]{Prockup2015}
Prockup, M.; Ehmann, A.F.; Gouyon, F.; Schmidt, E.M.; Celma, O.; Kim, Y.E.
\newblock {Modeling genre with the Music Genome Project: comparing
  human-labeled attributes and audio features}.
\newblock  Proc. 16th Int. Soc. Music Inform. Retrieval Conf.,  2015, pp.
  31--37.

\bibitem[Kim and Whitman(2002)]{Kim2002}
Kim, Y.E.; Whitman, B.
\newblock {Singer identification in popular music recordings using voice coding
  features}.
\newblock  Proc. 3rd Int. Conf. Music Inform. Retrieval,  2002, pp. 17--23.

\bibitem[Skowronek \em{et~al.}(2006)Skowronek, McKinney, and van~de
  Par]{Skowronek2006}
Skowronek, J.; McKinney, M.F.; van~de Par, S.
\newblock {Ground truth for automatic music mood classification.}
\newblock  Proc. 7th Int. Conf. Music Inform. Retrieval,  2006, pp. 395--396.

\bibitem[Sturm(2013)]{Sturm2013}
Sturm, B.L.
\newblock {The GTZAN dataset: its contents, its faults, their effects on
  evaluation, and its future use}.
\newblock {\em arXiv} {\bf 2013}, pp. 1--29,
  \href{http://xxx.lanl.gov/abs/1306.1461}{{\normalfont [1306.1461]}}.

\bibitem[Sturm(2015)]{Sturm2015}
Sturm, B.L.
\newblock {Faults in the latin music database and with its use}.
\newblock  Proc. Late Breaking Demo 16th Int. Soc. Music Inform. Retrieval
  Conf.,  2015, pp. 1--2.

\bibitem[Pachet and Roy(1999)]{Pachet1999}
Pachet, F.; Roy, P.
\newblock {Automatic generation of music programs}.
\newblock  Proc. 5th Int. Conf. Constraint Prog.; {Joxan Jaffar}., Ed.,  1999,
  pp. 331--345.

\bibitem[Eck \em{et~al.}(2007)Eck, Lamere, Bertin-Mahieux, and Green]{Eck2007}
Eck, D.; Lamere, P.; Bertin-Mahieux, T.; Green, S.
\newblock {Automatic generation of social tags for music recommendation}.
\newblock  Proc. 21st Conf. Adv. Neur. Inform. Process. Syst.,  2007, pp.
  385--392.

\bibitem[Li \em{et~al.}(2007)Li, Myaeng, and Kim]{Li2007}
Li, Q.; Myaeng, S.H.; Kim, B.M.
\newblock {A probabilistic music recommender considering user opinions and
  audio features}.
\newblock {\em Inform. Process. Manag.} {\bf 2007}, {\em 43},~473--487.

\bibitem[Schafer \em{et~al.}(2007)Schafer, Frankowski, Herlocker, and
  Sen]{Schafer2007}
Schafer, B.J.; Frankowski, D.; Herlocker, J.; Sen, S.
\newblock {Collaborative filtering recommender systems}. In {\em Adapt. Web}, 1
  ed.;  Brusilovski, P.; Kobsa, A.; Nejdl, W., Eds.; Springer-Verlag Berlin
  Heidelberg,  2007; chapter~9, pp. 291--324.

\bibitem[Schl{\"{u}}ter and Grill(2015)]{Schluter2015}
Schl{\"{u}}ter, J.; Grill, T.
\newblock {Exploring data augmentation for improved singing voice detection
  with neural networks}.
\newblock  Proc. 16th Int. Soc. Music Inform. Retrieval Conf.,  2015, pp.
  121--126.

\bibitem[Logan(2002)]{Logan2002}
Logan, B.
\newblock {Content-based playlist generation: exploratory experiments}.
\newblock  Proc. 3rd Int. Conf. Music Inform. Retrieval,  2002, pp. 6--7.

\bibitem[Hoashi \em{et~al.}(2003)Hoashi, Matsumoto, and Inoue]{Hoashi2003}
Hoashi, K.; Matsumoto, K.; Inoue, N.
\newblock {Personalization of user profiles for content-based music retrieval
  based on relevance feedback}.
\newblock  Proc. 11th ACM Int. Conf. Multimedia,  2003, pp. 110--119.

\bibitem[Celma \em{et~al.}(2005)Celma, Ram{\'{i}}rez, and Herrera]{Celma2005}
Celma, {\`{O}}.; Ram{\'{i}}rez, M.; Herrera, P.
\newblock {Foafing the music: a music recommendation system based on RSS feeds
  and user preferences}.
\newblock  Proc. 6th Int. Conf. Music Inform. Retrieval,  2005, pp. 457--464.

\bibitem[Sordo \em{et~al.}(2007)Sordo, Laurier, and Celma]{Sordo2007}
Sordo, M.; Laurier, C.; Celma, {\`{O}}.
\newblock {Annotating music collections: how content-based similarity helps to
  propagate labels}.
\newblock  Proc. 8th Int. Conf. Music Inform. Retrieval,  2007, pp. 531--534.

\bibitem[Tingle \em{et~al.}(2010)Tingle, Kim, and Turnbull]{Tingle2010}
Tingle, D.; Kim, Y.E.; Turnbull, D.
\newblock {Exploring automatic music annotation with "acoustically-objective"
  tags}.
\newblock  Proc. 11th ACM Int. Conf. Multimedia Inform. Retrieval,  2010, pp.
  55--62.

\bibitem[Bu \em{et~al.}(2010)Bu, Tan, Chen, Wang, Wu, Zhang, and He]{Bu2010}
Bu, J.; Tan, S.; Chen, C.; Wang, C.; Wu, H.; Zhang, L.; He, X.
\newblock {Music recommendation by unified hypergraph: combining social media
  information and music content}.
\newblock  Proc. 18th ACM Int. Conf. Multimedia,  2010, pp. 391--400.

\bibitem[Hsu \em{et~al.}(2016)Hsu, Lin, and Chi]{Hsu2016}
Hsu, K.C.; Lin, C.S.; Chi, T.S.
\newblock {Sparse coding based music genre classification using
  spectro-temporal modulations}.
\newblock  Proc. 17th Int. Soc. Music Inform. Retrieval Conf.,  2016, pp.
  744--750.

\bibitem[Jeong and Lee(2016)]{Jeong2016}
Jeong, I.Y.; Lee, K.
\newblock {Learning temporal features using a deep neural network and its
  application to music genre classification}.
\newblock  Proc. 17th Int. Soc. Music Inform. Retrieval Conf.,  2016, pp.
  434--440.

\bibitem[Lu \em{et~al.}(2016)Lu, Wu, Lu, and Lerch]{Lu2016}
Lu, Y.C.; Wu, C.W.; Lu, C.T.; Lerch, A.
\newblock {Automatic outlier detection in music genre datasets}.
\newblock  Proc. 17th Int. Soc. Music Inform. Retrieval Conf.,  2016, pp.
  101--107.

\bibitem[Oramas \em{et~al.}(2016)Oramas, Espinosa-Anke, Lawlor, Serra, and
  Saggion]{Oramas2016}
Oramas, S.; Espinosa-Anke, L.; Lawlor, A.; Serra, X.; Saggion, H.
\newblock {Exploring customer reviews for music genre classification and
  evolutionary studies}.
\newblock  Proc. 17th Int. Soc. Music Inform. Retrieval Conf.,  2016, pp.
  150--156.

\bibitem[Sturm(2016)]{Sturm2016}
Sturm, B.L.
\newblock {Revisiting priorities: improving MIR evaluation practices}.
\newblock  Proc. 17th Int. Soc. Music Inform. Retrieval Conf.,  2016, pp.
  488--494.

\bibitem[Wiggins(2009)]{Wiggins2009}
Wiggins, G.A.
\newblock {Semantic gap?? Schemantic schmap!! Methodological considerations in
  the scientific study of music}.
\newblock  Proc. 11th IEEE Int. Symp. Multimedia,  2009, pp. 477--482.

\bibitem[Chau \em{et~al.}(2013)Chau, Ho, Ho, and Yao]{Chau2013}
Chau, P.Y.K.; Ho, S.Y.; Ho, K.K.W.; Yao, Y.
\newblock {Examining the effects of malfunctioning personalized services on
  online users' distrust and behaviors}.
\newblock {\em Decision Support Systems} {\bf 2013}, {\em 56},~180--191.

\bibitem[Yin \em{et~al.}(2010)Yin, Bond, and Zhang]{Yin2010}
Yin, D.; Bond, S.D.; Zhang, H.
\newblock {Are bad reviews always stronger than good? Asymmetric negativity
  bias in the formation of online consumer trust}.
\newblock  Proc. 31st Int. Conf. Inform. Syst.,  2010, pp. 1--18.

\bibitem[Gouyon \em{et~al.}(2014)Gouyon, Sturm, Oliveira, Hespanhol, and
  Langlois]{Gouyon2014}
Gouyon, F.; Sturm, B.L.; Oliveira, J.L.; Hespanhol, N.; Langlois, T.
\newblock {On evaluation validity in music autotagging}.
\newblock {\em arXiv} {\bf 2014},
  \href{http://xxx.lanl.gov/abs/1410.0001}{{\normalfont [1410.0001]}}.

\bibitem[Rosenblatt(2015)]{Rosenblatt2015}
Rosenblatt, D.
\newblock {Music Listening as Therapy}.
\newblock PhD thesis, Univ. Loma Linda, CA, USA,  2015.

\bibitem[Su{\'{a}}rez \em{et~al.}(2016)Su{\'{a}}rez, Elangovan, and
  Au]{Suarez2016}
Su{\'{a}}rez, L.; Elangovan, S.; Au, A.
\newblock {Cross-sectional study on the relationship between music training and
  working memory in adults}.
\newblock {\em Australian J. Psych.} {\bf 2016}, {\em 68},~38--46.

\bibitem[Zhao and Kuhl(2016)]{Zhao2016}
Zhao, T.C.; Kuhl, P.K.
\newblock {Musical intervention enhances infants' neural processing of temporal
  structure in music and speech}.
\newblock {\em Proc. Natl. Acad. Sci. USA} {\bf 2016}, {\em 113},~5212--5217.

\bibitem[Rao \em{et~al.}(2009)Rao, Ramakrishnan, and Rao]{Rao2009}
Rao, V.; Ramakrishnan, S.; Rao, P.
\newblock {Singing voice detection in polyphonic music using predominant
  pitch}.
\newblock  Proc. 10th Annu. Conf. Inter. Speech Comm. Assoc.,  2009, pp.
  1131--1134.

\bibitem[Panteli \em{et~al.}(2017)Panteli, Bittner, Bello, and
  Dixon]{Panteli2017}
Panteli, M.; Bittner, R.; Bello, J.P.; Dixon, S.
\newblock {Towards the characterization of singing styles in world music}.
\newblock  Proc. IEEE Int. Conf. Acoust. Speech Signal Process.,  2017, pp.
  636--640.

\bibitem[Ghosal \em{et~al.}(2013)Ghosal, Chakraborty, Dhara, and
  Saha]{Ghosal2013}
Ghosal, A.; Chakraborty, R.; Dhara, B.C.; Saha, S.K.
\newblock {A hierarchical approach for speech-instrumental-song
  classification}.
\newblock {\em SpringerPlus} {\bf 2013}, {\em 2},~1--11.

\bibitem[Bayle \em{et~al.}(2016)Bayle, Hanna, and Robine]{Bayle2016}
Bayle, Y.; Hanna, P.; Robine, M.
\newblock {Classification {\`{a}} grande {\'{e}}chelle de morceaux de musique
  en fonction de la pr{\'{e}}sence de chant}.
\newblock  {Journées d'Informatique Musicale},  2016, pp. 144--152.

\bibitem[Bogdanov \em{et~al.}(2016)Bogdanov, Porter, Herrera, and
  Serra]{Bogdanov2016}
Bogdanov, D.; Porter, A.; Herrera, P.; Serra, X.
\newblock {Cross-collection evaluation for music classification tasks}.
\newblock  Proc. 17th Int. Soc. Music Inform. Retrieval Conf.,  2016, pp.
  379--385.

\bibitem[Lehner \em{et~al.}(2014)Lehner, Widmer, and Sonnleitner]{Lehner2014}
Lehner, B.; Widmer, G.; Sonnleitner, R.
\newblock {On the reduction of false positives in singing voice detection}.
\newblock  Proc. IEEE Int. Conf. Acoust. Speech Signal Process.,  2014, pp.
  7480--7484.

\bibitem[Casey \em{et~al.}(2008)Casey, Veltkamp, Goto, Leman, Rhodes, and
  Slaney]{Casey2008}
Casey, M.A.; Veltkamp, R.; Goto, M.; Leman, M.; Rhodes, C.; Slaney, M.
\newblock {Content-based music information retrieval: Current directions and
  future challenges}.
\newblock {\em Proc. IEEE} {\bf 2008}, {\em 96},~668--696.

\bibitem[Bayle \em{et~al.}(2017)Bayle, Hanna, and Robine]{Bayle2017}
Bayle, Y.; Hanna, P.; Robine, M.
\newblock {SATIN: A Persistent Musical Database for Music Information
  Retrieval}.
\newblock  Proc. 15th Int. Works. Content-Based Multimedia Indexing,  2017, pp.
  1--5.

\bibitem[Ramona \em{et~al.}(2008)Ramona, Richard, and David]{Ramona2008}
Ramona, M.; Richard, G.; David, B.
\newblock {Vocal detection in music with support vector machines}.
\newblock  Proc. IEEE Int. Conf. Acoust. Speech Signal Process.,  2008, pp.
  1885--1888.

\bibitem[Bittner \em{et~al.}(2014)Bittner, Salamon, Tierney, Mauch, Cannam, and
  Bello]{Bittner2014}
Bittner, R.M.; Salamon, J.; Tierney, M.; Mauch, M.; Cannam, C.; Bello, J.P.
\newblock {MedleyDB: a multitrack dataset for annotation-intensive MIR
  research}.
\newblock  Proc. 15th Int. Soc. Music Inform. Retrieval Conf.,  2014, pp.
  155--160.

\bibitem[Liutkus \em{et~al.}(2014)Liutkus, Fitzgerald, Rafii, Pardo, and
  Daudet]{Liutkus2014}
Liutkus, A.; Fitzgerald, D.; Rafii, Z.; Pardo, B.; Daudet, L.
\newblock {Kernel additive models for source separation}.
\newblock {\em {IEEE} Trans. Signal Process.} {\bf 2014}, {\em 62},~4298--4310.

\bibitem[Schl{\"{u}}ter(2016)]{Schluter2016}
Schl{\"{u}}ter, J.
\newblock {Learning to pinpoint singing voice from weakly labeled examples}.
\newblock  Proc. 17th Int. Soc. Music Inform. Retrieval Conf.,  2016, pp.
  44--50.

\bibitem[Hespanhol(2013)]{Hespanhol2013}
Hespanhol, N.
\newblock {Using Autotagging for Classification of Vocals in Music Signals}.
\newblock PhD thesis, Univ. Porto, Portugal,  2013.

\bibitem[Zhang and Kuo(2013)]{Zhang2013}
Zhang, T.; Kuo, C.C.J.
\newblock {\em {Content-Based Audio Classification and Retrieval for
  Audiovisual Data Parsing}}; Springer Science {\&} Business Media,  2013; p.
  136.

\bibitem[Sturm(2014)]{Sturm2014b}
Sturm, B.L.
\newblock The state of the art ten years after a state of the art: Future
  research in music information retrieval.
\newblock {\em Journal of New Music Research} {\bf 2014}, {\em 43},~147--172.

\bibitem[Fischler and Bolles(1981)]{Fischler1981}
Fischler, M.A.; Bolles, R.C.
\newblock {Random sample consensus: a paradigm for model fitting with
  application to image analysis and automated cartography}.
\newblock {\em Commun. ACM} {\bf 1981}, {\em 24},~381--395.

\bibitem[Ness \em{et~al.}(2009)Ness, Theocharis, Tzanetakis, and
  Martins]{Ness2009}
Ness, S.R.; Theocharis, A.; Tzanetakis, G.; Martins, L.G.
\newblock {Improving automatic music tag annotation using stacked
  generalization of probabilistic SVM outputs}.
\newblock  Proc. 17th ACM Int. Conf. Multimedia,  2009, pp. 705--708.

\bibitem[Langlois and Marques(2009)]{Langlois2009}
Langlois, T.; Marques, G.
\newblock {A music classification method based on timbral features}.
\newblock  Proc. 10th Int. Soc. Music Inform. Retrieval Conf.,  2009, pp.
  81--86.

\bibitem[Panagakis \em{et~al.}(2009)Panagakis, Kotropoulos, and
  Arce]{Panagakis2009}
Panagakis, Y.; Kotropoulos, C.; Arce, G.R.
\newblock {Music genre classification via sparse representations of auditory
  temporal modulations}.
\newblock  Proc. 17th European Signal Process. Conf.,  2009, pp. 1--5.

\bibitem[Wright \em{et~al.}(2009)Wright, Yang, Ganesh, Sastry, and
  Ma]{Wright2009}
Wright, J.; Yang, A.Y.; Ganesh, A.; Sastry, S.S.; Ma, Y.
\newblock {Robust face recognition via sparse representation}.
\newblock {\em {IEEE} Trans. Pattern Anal. Mach. Intell.} {\bf 2009}, {\em
  31},~210--227.

\bibitem[Sturm(2012)]{Sturm2012}
Sturm, B.L.
\newblock {Two systems for automatic music genre recognition: what are they
  really recognizing?}
\newblock  Proc. 2nd Int. ACM Works. Music Inf. Retrieval User-Centered
  Multimodal Strat.,  2012, pp. 69--74.

\bibitem[Sturm and Noorzad(2012)]{Sturm2012a}
Sturm, B.L.; Noorzad, P.
\newblock {On automatic music genre recognition by sparse representation
  classification using auditory temporal modulations}.
\newblock  Proc. 9th Int. Symp. Comp. Music Model. Retrieval,  2012, pp.
  379--394.

\bibitem[Pedregosa \em{et~al.}(2011)Pedregosa, Varoquaux, Gramfort, Michel,
  Thirion, Grisel, Blondel, Prettenhofer, Weiss, Dubourg, Vanderplas, Passos,
  Cournapeau, Brucher, Perrot, and Duchesnay]{scikit-learn}
Pedregosa, F.; Varoquaux, G.; Gramfort, A.; Michel, V.; Thirion, B.; Grisel,
  O.; Blondel, M.; Prettenhofer, P.; Weiss, R.; Dubourg, V.; Vanderplas, J.;
  Passos, A.; Cournapeau, D.; Brucher, M.; Perrot, M.; Duchesnay, {\'{E}}.
\newblock {Scikit-learn: machine learning in Python}.
\newblock {\em J. Mach. Learning Res.} {\bf 2011}, {\em 12},~2825--2830.

\bibitem[Livshin and Rodet(2003)]{Livshin2003}
Livshin, A.; Rodet, X.
\newblock {The importance of cross database evaluation in sound
  classification}.
\newblock  Proc. 4th Int. Conf. Music Inform. Retrieval,  2003, pp. 1--2.

\bibitem[Guaus(2009)]{Guaus2009}
Guaus, E.
\newblock {Audio Content Processing for Automatic Music Genre Classification:
  Descriptors, Databases, and Classifiers}.
\newblock PhD thesis, Univ. Pompeu Fabra, Barcelona, Spain,  2009.

\bibitem[Ng(1997)]{Ng1997}
Ng, A.Y.
\newblock {Preventing "overfitting" of cross-validation data}.
\newblock  Proc. 14th Int. Conf. Mach. Learning,  1997, pp. 245--253.

\bibitem[Herrera \em{et~al.}(2003)Herrera, Dehamel, and Gouyon]{Herrera2003}
Herrera, P.; Dehamel, A.; Gouyon, F.
\newblock {Automatic labeling of unpitched percussion sounds}.
\newblock  Proc. 114th Audio Eng. Soc. Conv.,  2003, pp. 1--14.

\bibitem[Bogdanov \em{et~al.}(2011)Bogdanov, Serr{\`{a}}, Wack, Herrera, and
  Serra]{Bogdanov2011}
Bogdanov, D.; Serr{\`{a}}, J.; Wack, N.; Herrera, P.; Serra, X.
\newblock {Unifying low-level and high-level music similarity measures}.
\newblock {\em {IEEE} Trans. Multimedia} {\bf 2011}, {\em 13},~687--701.

\bibitem[Chud{\'{a}}{\v{c}}ek \em{et~al.}(2009)Chud{\'{a}}{\v{c}}ek,
  Georgoulas, Lhotsk{\'{a}}, Stylios, Petr{\'{i}}k, and
  {\v{C}}epek]{Chudacek2009}
Chud{\'{a}}{\v{c}}ek, V.; Georgoulas, G.; Lhotsk{\'{a}}, L.; Stylios, C.;
  Petr{\'{i}}k, M.; {\v{C}}epek, M.
\newblock {Examining cross-database global training to evaluate five different
  methods for ventricular beat classification}.
\newblock {\em J. Physio. Measurement} {\bf 2009}, {\em 30},~661--677.

\bibitem[Bekios-Calfa \em{et~al.}(2011)Bekios-Calfa, Buenaposada, and
  Baumela]{Bekios-Calfa2011}
Bekios-Calfa, J.; Buenaposada, J.M.; Baumela, L.
\newblock {Revisiting linear discriminant techniques in gender recognition}.
\newblock {\em {IEEE} Trans. Pattern Anal. Mach. Intell.} {\bf 2011}, {\em
  33},~858--864.

\bibitem[Llamedo \em{et~al.}(2012)Llamedo, Khawaja, and Martinez]{Llamedo2012}
Llamedo, M.; Khawaja, A.; Martinez, J.P.
\newblock {Cross-database evaluation of a multilead heartbeat classifier}.
\newblock {\em {IEEE} Trans. Inf. Technol. Biomed.} {\bf 2012}, {\em
  16},~658--664.

\bibitem[Erdoğmuş \em{et~al.}(2014)Erdoğmuş, Vanoni, and
  Marcel]{Erdogmus2014}
Erdoğmuş, N.; Vanoni, M.; Marcel, S.
\newblock {Within- and cross- database evaluations for face gender
  classification via BeFIT protocols}.
\newblock  Proc. 16th IEEE Int. Works. Multimedia Signal Process.,  2014, pp.
  1--6.

\bibitem[Fern{\'{a}}ndez \em{et~al.}(2015)Fern{\'{a}}ndez, Huerta, and
  Prati]{Fernandez2015}
Fern{\'{a}}ndez, C.; Huerta, I.; Prati, A.
\newblock {A comparative evaluation of regression learning algorithms for
  facial age estimation}. In {\em Face and Facial Expression Recognition from
  Real World Videos}; Ji, Q.; Moeslund, T.; Hua, G.; Nasrollahi, K., Eds.;
  Springer, Cham,  2015; pp. 133--144.

\bibitem[Sturm(2014)]{Sturm2014a}
Sturm, B.L.
\newblock {A simple method to determine if a music information retrieval system
  is a "Horse"}.
\newblock {\em {IEEE} Trans. Multimedia} {\bf 2014}, {\em 16},~1636--1644.

\bibitem[Foucard \em{et~al.}(2012)Foucard, Essid, Lagrange, and
  Richard]{Foucard2012}
Foucard, R.; Essid, S.; Lagrange, M.; Richard, G.
\newblock {{\'{E}}tiquetage automatique de musique : une approche de boosting
  r{\'{e}}gressif bas{\'{e}}e sur une fusion souple d'annotateurs}.
\newblock  Proc. 15th Conf. Compression Representation Signaux Audiovisuels,
  2012, pp. 169--173.

\bibitem[Sturm \em{et~al.}(2014)Sturm, Bardeli, Langlois, and Emiya]{Sturm2014}
Sturm, B.L.; Bardeli, R.; Langlois, T.; Emiya, V.
\newblock {Formalizing the problem of music description}.
\newblock  Proc. 15th Int. Soc. Music Inform. Retrieval Conf.,  2014, pp.
  89--94.

\bibitem[Krizhevsky \em{et~al.}(2012)Krizhevsky, Sutskever, and
  Hinton]{Krizhevsky2012}
Krizhevsky, A.; Sutskever, I.; Hinton, G.E.
\newblock {ImageNet Classification with Deep Convolutional Neural Networks}. In
  {\em Proc. 25th Conf. Advances Neur. Inform. Proc. Syst.}; Pereira, F.;
  Burges, C.J.C.; Bottou, L.; Weinberger, K.Q., Eds.; Curran Associates, Inc.,
  2012; pp. 1097--1105.

\bibitem[Chawla(2009)]{Chawla2009}
Chawla, N.V.
\newblock {Data mining for imbalanced datasets: An overview}. In {\em {Data
  mining and knowledge discovery handbook}}; {Springer US}., Ed.;  2009; pp.
  875--886.

\bibitem[Wong \em{et~al.}(2016)Wong, Gatt, Stamatescu, and McDonnell]{Wong2016}
Wong, S.C.; Gatt, A.; Stamatescu, V.; McDonnell, M.D.
\newblock {Understanding Data Augmentation for Classification: When to Warp?}
\newblock  Proc. Int. Conf. Digital Image Comp. Tech. App.,  2016, pp. 1--6.

\bibitem[Bergstra \em{et~al.}(2006)Bergstra, Casagrande, Erhan, Eck, and
  Kegl]{Bergstra2006}
Bergstra, J.; Casagrande, N.; Erhan, D.; Eck, D.; Kegl, B.
\newblock {Meta-Features and AdaBoost for Music Classification}.
\newblock {\em {Machine Learning}} {\bf 2006}, pp. 1--28.

\bibitem[Lukashevich \em{et~al.}(2007)Lukashevich, Gruhne, and
  Dittmar]{Lukashevich2007}
Lukashevich, H.; Gruhne, M.; Dittmar, C.
\newblock {Effective singing voice detection in popular music using arma
  filtering}.
\newblock  Proc. 10th Int. Works. Digital Audio Effects,  2007, pp. 165--168.

\bibitem[Regnier and Peeters(2009)]{Regnier2009}
Regnier, L.; Peeters, G.
\newblock {Singing voice detection in music tracks using direct voice vibrato
  detection}.
\newblock  Proc. IEEE Int. Conf. Acoust. Speech Signal Process.,  2009, pp.
  1685--1688.

\bibitem[Leglaive \em{et~al.}(2015)Leglaive, Hennequin, and
  Badeau]{Leglaive2015}
Leglaive, S.; Hennequin, R.; Badeau, R.
\newblock {Singing voice detection with deep recurrent neural networks}.
\newblock  Proc. 40th IEEE Int. Conf. Acoust. Speech Signal Process.,  2015,
  pp. 121--125.

\bibitem[Lehner \em{et~al.}(2015)Lehner, Widmer, and B{\"{o}}ck]{Lehner2015}
Lehner, B.; Widmer, G.; B{\"{o}}ck, S.
\newblock {A low-latency, real-time-capable singing voice detection method with
  LSTM recurrent neural networks}.
\newblock  Proc. 23rd European Signal Process. Conf.,  2015, pp. 21--25.

\bibitem[Nwe \em{et~al.}(2004)Nwe, Shenoy, and Wang]{Nwe2004}
Nwe, T.L.; Shenoy, A.; Wang, Y.
\newblock {Singing voice detection in popular music}.
\newblock  Proc. 12th Annu. ACM Int. Conf. Multimedia,  2004, pp. 324--327.

\bibitem[Kereliuk \em{et~al.}(2015)Kereliuk, Sturm, and Larsen]{Kereliuk2015}
Kereliuk, C.; Sturm, B.L.; Larsen, J.
\newblock {Deep learning and music adversaries}.
\newblock {\em {IEEE} Trans. Multimedia} {\bf 2015}, {\em 17},~2059--2071.

\bibitem[Lidy and Schindler(2016)]{Lidy2016}
Lidy, T.; Schindler, A.
\newblock {CQT-based convolutional neural networks for audio scene
  classification and Domestic Audio Tagging}.
\newblock  Proc. IEEE Audio Acoust. Signal Process. Challenge Works. Detect.
  Classif. Acoustic Scenes Events,  2016, pp. 60--64.

\bibitem[Pons \em{et~al.}(2016)Pons, Lidy, and Serra]{Pons2016}
Pons, J.; Lidy, T.; Serra, X.
\newblock {Experimenting with musically motivated convolutional neural
  networks}.
\newblock  Proc. 14th Int. Works. Content-Based Multimedia Indexing,  2016, pp.
  1--6.

\bibitem[Woods and Bowyer(1997)]{Woods1997}
Woods, K.; Bowyer, K.W.
\newblock Generating ROC curves for artificial neural networks.
\newblock {\em {IEEE} Trans. Med. Imag.} {\bf 1997}, {\em 16},~329--337.

\bibitem[Zhao \em{et~al.}(2011)Zhao, Jin, Yang, and Hoi]{Zhao2011}
Zhao, P.; Jin, R.; Yang, T.; Hoi, S.C.
\newblock Online AUC maximization.
\newblock  Proc. 28th Int. Conf. Mach. Learn.,  2011, pp. 233--240.

\bibitem[Miron \em{et~al.}(2017)Miron, Janer~Mestres, and
  G{\'o}mez~Guti{\'e}rrez]{Miron2017a}
Miron, M.; Janer~Mestres, J.; G{\'o}mez~Guti{\'e}rrez, E.
\newblock Generating data to train convolutional neural networks for classical
  music source separation.
\newblock  Proc. 14th Sound Music Comp. Conf.,  2017, pp. 227--234.

\bibitem[Oramas \em{et~al.}(2017)Oramas, Nieto, Sordo, and Serra]{Oramas2017a}
Oramas, S.; Nieto, O.; Sordo, M.; Serra, X.
\newblock A deep multimodal approach for cold-start music recommendation.
\newblock  Proc. 2nd Work. Deep Learn. Rec. Syst.,  2017, pp. 32--37.

\end{thebibliography}


\end{document}